\documentclass{sig-alternate-05-2015}
\newcommand*{\refname}{Bibliography}

\usepackage[normalem]{ulem}

\newcommand{\mt}[1]{\texttt{#1}}
\usepackage[all]{nowidow}
\usepackage{color}
\usepackage{setspace}
\usepackage{subfig}
\usepackage{flushend}

\usepackage{fancyhdr}
  \usepackage{cite}
   \usepackage{graphicx}
\usepackage{amsmath}
\usepackage{booktabs}

\begin{document}

\setlength{\parskip}{0.5em plus 0.5em minus 0.2em}

%\CopyrightYear{2016}
%\setcopyright{acmcopyright}
%\conferenceinfo{PACT '16,}{September 11-15, 2016, Haifa, Israel}
%\isbn{978-1-4503-4121-9/16/09}\acmPrice{\$15.00}
%\doi{http://dx.doi.org/10.1145/2967938.2967966}

\title{Vectorization of Multibyte Floating Point Data Formats}


\numberofauthors{2}

\author{
\alignauthor
Andrew Anderson\\
\affaddr{Lero, Trinity College Dublin}\\
\email{aanderso@cs.tcd.ie}
\alignauthor
David Gregg\\
\affaddr{Lero, Trinity College Dublin}\\
\email{dgregg@cs.tcd.ie}
}

\maketitle

\begin{abstract}

\noindent We propose a scheme for reduced-precision representation of floating
point data on a continuum between IEEE-754 floating point types. Our scheme
enables the use of lower precision formats for a reduction in storage space
requirements and data transfer volume. We describe how our scheme can be
accelerated using existing hardware vector units on a general-purpose
processor (GPP). Exploiting native vector hardware allows us to support reduced
precision floating point with low overhead. We demonstrate that supporting
reduced precision in the compiler as opposed to using a library approach can
yield a low overhead solution for GPPs.

%Using our approach, we achieve
%improvements of up to $6\times$ and $4.5\times$ fewer last-level cache misses
%on programs from BLAS Level 2 and BLAS Level 3, respectively.

%Our approach does not
%require custom hardware, or native support for non-standard floating point.

\end{abstract}





\section{Motivation}\label{sec:motivation}

It has long been recognized that different applications and algorithms need
different amounts of floating point precision to achieve accurate results
\cite{tong2000reducing,buttari2006exploiting,rubio2013precimonious}. For
example, 64-bit double precision floating point is often needed for scientific
applications, whereas 32-bit single precision is sufficient for most graphics
and computer vision applications.

Modern GPUs and embedded processors increasingly support 16-bit floating point
for applications that are highly tolerant of approximation such as speech
recognition \cite{dixon2009fast}.

In some contexts it is possible to customize the level of precision precisely
to the application. For example, field-programmable gate arrays (FPGAs) can be
used to implement customized floating point with bit-level control over the
size of the mantissa and exponent \cite{de2009generating,tong2000reducing}.

More recently, Schkufza et al. \cite{schkufza2014stochastic} have shown how
superoptimization can be used to generate iterative floating point algorithms
that guarantee a given level of accuracy. For example, an exponential
function may return a 64-bit floating point value in which at least the first,
say, 40 bits of the mantissa are guaranteed to be accurate.

A problem with customizing floating point precision on general-purpose
processors (GPPs) is that most support only two standard types: IEEE-754 single
precision (or \mt{binary32}) and double precision (\mt{binary64}). If an
operation needs more precision than \mt{binary32} and less than
\mt{binary64}, the developer has no choice but to use \mt{binary64}.

%Try to get across: it would be nice if we could do custom precision on gpps
%and not have to use FPGAs.

%\subsection*{Proposal}

In this paper we propose a compiler-based mechanism for supporting several
non-standard \emph{multibyte} floating point memory formats, such as 24-, 40-,
or 48-bit floating point. By \emph{multibyte} we mean that these formats differ
in length from standard IEEE-754 formats by multiples of 8 bits.

By using these types, a developer can reduce the precision of floating point
data in memory, resulting in reduced storage requirements for their data.

An increasingly important factor in the design of computing systems is energy
consumption. In embedded computing systems the energy consumed by data movement
can be much greater than the energy used to perform arithmetic
\cite{dally2008efficient}.

According to Gustafson~\cite{gustafson}, a typical 64-bit multiply-add
operation costs around 200pJ, while reading 64 bits from cache costs 800pJ and
a 64-bit main memory read costs 12000pJ. Reducing the amount of data movement
is therefore essential to reducing energy consumed by an application, and in
particular, reducing the number of costly last-level cache misses.

Customizing precision of floating point data to the needs of the application is
one way to reduce data movement. It is straightforward to implement a C/C++
data type representing, for example, 24-bit floating point values
(Figure~\ref{fig:simple-flyte}).

However, we have found that the performance of such code can be extraordinarily
poor when operating on arrays of data. We therefore propose a technique to
generate vectorized code that performs a number of adjacent reads or writes
together, along with the required packing or unpacking.

%Our technique locates conversion directly in the loop body, which
%can result in significant instruction-level parallelism between conversion and
%computation on modern GPPs, which tend to be out-of-order superscalar designs.
%We make the following contributions:

\subsubsection*{Contributions}

\begin{itemize}

\item We present a practical representation of multibyte floating
  point values that can easily be converted to and from the next
  largest natively-supported type.

\item We propose a compiler vectorization scheme for packing and
  unpacking our floating point types, enabling their use in vectorized floating
  point computations.

%\item We provide a number of software schemes for rounding
  %standard-precision floating point values to our reduced-precision
  %types with various trade-offs in accuracy and efficiency.

%\item We show how to deal correctly with special values such as
  %sub-normal numbers, inifinites, and non-signalling NaN.

\item We demonstrate experimentally that our techniques provide a low-overhead
way to support customized floating point types on general-purpose processors.

\end{itemize}

\section{Customizing Floating Point}
%\section{IEEE-754 Floating Point}
\label{sec:ieee-754-intro}

The IEEE-754 2008 standard~\cite{zuras2008ieee} defines a number of finite
binary representations of real numbers at different resolutions (16, 32, and 64
bits, among others). Each format encodes floating-point values using three
binary integers: \emph{sign}, \emph{exponent}, and \emph{mantissa}, which
specify a point on the real number line following a general formula:
\begin{align*}
v = (-1)^{s} \times (1 + \sum_{i=1}^{M}(m_{i}2^{-i})) \times 2^{e-bias}
\end{align*}

\noindent $v$ is the real number obtained by the evaluation of the formula for
a floating-point value with sign $s$, mantissa $m$ of length $M$ bits, and
exponent $e$. The value $bias$ is an integer constant which differs for each
format. Different formats (\mt{binary32}, \mt{binary64}, etc.) use different
numbers of bits for the exponent and mantissa components.

The exponent determines a power of two by which the rest of the number is
multiplied, while the mantissa represents a fractional quantity in the interval
$[1,2)$ obtained by summing successively smaller powers of two, starting from
$2^{-1}$ and continuing up to the length of the mantissa. If the $i$th mantissa
bit is set, the corresponding power of two is present in the sum. For
normalized numbers (those with a nonzero exponent), the leading 1 is not
explicitly stored in the mantissa, but is implied.

The structure of the IEEE-754 binary encoding means that a change in an
exponent bit strongly influences the resulting value, while a change in a
mantissa bit has less influence. Furthermore, a change in any bit of the
mantissa has exponentially greater effect on the resulting value than a change
in the next-least-significant bit.

These observations lead naturally to a scheme for representing values at
precisions between those specified by the IEEE-754 standard: varying the number
of low-order mantissa bits. Previous proposals based around this concept have
typically made use of either customized floating point hardware support via
FPGA~\cite{de2009generating} or high-level data reorganization in massively
parallel systems~\cite{jenkins2012byte}.

In this paper, we demonstrate how reduced precision floating point
representations can be used on general-purpose processors without requiring any
special hardware support.

The rest of our paper is organized as follows: first, in
Section~\ref{sec:scheme}, we discuss our scheme for representing floating point
numbers with reduced precision in memory, and a straightforward implementation
of the scheme in scalar code using a library of datatypes implementing
different memory representations of floating point numbers with different
precision.

Next, in Section~\ref{sec:rw}, we discuss aspects of contemporary GPP
architecture which confound the straightforward library approach, and propose a
vectorized, compiler-accelerated approach. Section~\ref{sec:rounding} discusses
\emph{rounding}, an important aspect of implementing floating point.
Section~\ref{sec:eval} presents an experimental evaluation of both library and
compiler-accelerated schemes on a recent general-purpose processor.
Section~\ref{sec:related} discusses related work, and
Section~\ref{sec:conclusion} concludes.

%Our 24-bit floating point thought experiment (Figure~\ref{fig:float-example})
%shows one of these short representations. By truncating the unused low 8 bits
%of the mantissa, we can store reduced precision numbers in a compact
%representation and free up 25\% of the memory which would have been used if the
%results were stored using the full 32 bits of the IEEE-754 single-precision
%format.

%Figure~\ref{fig:nsfloat} displays some possible reduced-precision
%representations for IEEE-754 32-bit floating point.

%\begin{figure}[h]
%\centering\includegraphics[scale=0.8]{figures/nsfloat-diagram}
%\caption{Layout of alternative multibyte floating point formats}
%\label{fig:nsfloat}
%\end{figure}

% A key feature of the representations defined in IEEE-754 is that
% their width is \emph{fixed}. This can lead to scenarios where the
% full precision of a standard floating-point type is not required in
% the normal operation of the application.

\section{Reduced Precision on GPPs}
\label{sec:scheme}

It is possible to \textit{emulate} floating point operations in software using
integer instructions. However, each floating point operation requires many
integer instructions so emulation is slow.

In particular, the IEEE-754 standard has several special values and ranges such
as not-a-number (NaN), positive and negative zero, infinities, and sub-normal
numbers, each of which adds to the complexity of software emulation
\cite{sidwell2006improving}.

Modern processors provide hardware floating point units which dramatically
reduce the time and energy required for floating point computations. However,
these units normally support just two floating point types, typically
\mt{binary32} (\mt{float}) and \mt{binary64} (\mt{double}).

%The obvious way to support floating point types with non-standard sizes is to
%use software emulation, which is slow. In this paper we propose an alternative
%approach. Rather than using pure emulation, we store our multiple-byte floating
%point values in memory. When we operate on the non-standard values we first
%convert them to the next largest floating point type with hardware support, and
%use the existing hardware units to perform the computation. We then convert the
%result back down to our non-standard type for storage.

%A straightforward scheme for the use of reduced-precision storage formats on
%machines which only support computation on \mt{binary32} or \mt{binary64}
%values is to convert values from reduced-precision storage format into the
%appropriate IEEE-754 format when they are loaded. These values can then be used
%in computation as normal, producing results in a natively supported IEEE-754
%format. If values need to be stored in reduced-precision format, they can be
%converted while writing the data, storing it in the desired format in memory.
%The principal difference between the reading and writing stages are that the
%datatype widens on reads, and narrows on writes.

\subsection{Our Approach}
\label{sec:approach}

We propose a set of non-standard \textbf{fl}oating point multib\textbf{yte}
types that we refer to as \textit{flyte}s. Rather than emulating computation on
\textit{flyte}s in software, we convert them to/from the next largest natively
supported type. Thus, a 24-bit \textit{flyte} (\mt{flyte24}) is converted to
\mt{binary32} before computation, and the \mt{binary32} result is converted
back to \mt{flyte24} afterwards.

We need to solve two problems: (1) efficiently
loading and storing non-standard data sizes, such as 24-bit data, where there
is no hardware support for such operations, and (2) quickly converting between
built-in floating point types and our non-standard types.

In general, converting between floating point formats has many special cases.
In particular, converting between formats with different-sized exponents may
cause numbers to overflow to infinity or underflow to/from sub-normal numbers.
Dealing correctly with these cases in software is complicated and slow.

Our solution to the problem is that in all our \textit{flyte} types, the size
of the exponent is equal to the size of the exponent of the \textbf{next
largest built-in type}. For example, in our \mt{flyte16} and \mt{flyte24}
types, the exponent has eight bits, just like \mt{binary32}. This dramatically
reduces the complexity of conversions.

Efficiently supporting non-standard floating point types using this approach
creates two types of problems. The first is supporting the loading, storing,
and conversion of reduced precision types with acceptably low overhead. This is
the topic of the majority of this paper.

The second type of problem is that performing computation in one floating point
type and storing values in a less precise type introduces issues, such as
double-rounding, that complicate numerical analysis. We make every effort to be
clear about this latter group of problems but in most cases we do not have
comprehensive solutions. Double rounding in particular is a topic of extensive
study, and we refer the reader to the work of Boldo and
Melquiond~\cite{boldo2004double} for an in-depth discussion.

Our techniques are aimed squarely at problems where some approximation is
acceptable and the developer has a good understanding of exactly how much
precision is required. Our main contribution is to show \textit{how} to
implement multibyte floating point formats efficiently; the question of
\textit{whether} to use them in any particular algorithm depends on the
numerical properties of the algorithm.

\subsection{Simple Scalar Code Approach}
\label{sec:scalar-approach}

Figure \ref{fig:simple-flyte} shows a simple implementation of the \mt{flyte24}
type in C++. It relies on the \emph{bit field} facility in C/C++ to specify
that the \mt{num} field contains a 24-bit integer. It also uses the GCC
\texttt{packed} attribute to indicate to the compiler that arrays of the type
should be packed to exactly 24 bits, rather than padded to 32 bits. Figure
\ref{fig:simple-flyte} also shows a routine for converting from \mt{flyte24} to
\mt{float} (i.e. \mt{binary32}). The 24-bit pattern stored in the \mt{flyte24}
variable is scaled to 32 bits and padded with zeros. The resulting 32-bit
pattern is returned as a \mt{float}.

\begin{figure}[ht]
\begin{verbatim}
class flyte24 {
private:
  unsigned num:24;
public:
  operator float() {
    u32 temp = num << 8;
    return(cast_u32_to_f32(temp));
  };
  ...
} __attribute__((packed));
\end{verbatim}
\vspace{-9pt}
\caption{Simple implementation of \mt{flyte24} in C++}
\label{fig:simple-flyte}
\end{figure}

\noindent The code that is sketched in Fig. \ref{fig:simple-flyte} can be used
to implement programs with arrays of \mt{flyte24} values, but it is very slow.
Figure \ref{fig:realworldlib} shows a comparison of the execution time of
several BLAS kernels using \mt{flyte24} and other \mt{flyte} and IEEE-754
types.

The order of magnitude difference in execution time is the result of (1) the
cost of converting between \mt{flyte24} and \mt{binary32} before and after each
computation; and (2) the cost of loading and storing individual 24-bit values.

In particular, storing data to successive elements of packed \mt{flyte24}
arrays can result in sequences of overlapping 3-byte aligned stores. Load/store
hardware in GPPs is not designed to deal with such operations, which results in
extraordinarily slow execution.

Table~\ref{table:formats} summarizes our
proposed set of \emph{multibyte} formats for floating-point values which
preserve the sign and exponent fields of the corresponding IEEE-754
representations.

%The \textit{flyte} formats provide storage options along a continuum between
%the IEEE-754 \mt{binary32} and \mt{binary64} types.

\begin{table}[h]
  \caption{\textit{flyte} storage formats for IEEE-754 types.}
  \centering
  \begin{tabular}{ccccc}
    \hfill & \hfill & \multicolumn{3}{c}{\textit{flyte} layout (bits)} \\
    \toprule
    IEEE-754 type & \textit{flyte} format & Sign & Exp. & Mant. \\
    \midrule
    \mt{binary32} & 16-bit  & 1& 8& 7\\
    \mt{binary32} & 24-bit  & 1& 8& 15\\
    \mt{binary32} & 32-bit  & 1& 8& 23\\
    \midrule
    \mt{binary64} & 40-bit & 1& 11& 28\\
    \mt{binary64} & 48-bit & 1& 11& 36\\
    \mt{binary64} & 56-bit & 1& 11& 44\\
    \mt{binary64} & 64-bit & 1& 11& 52\\
    \bottomrule
  \end{tabular}
  \label{table:formats}
\end{table}

%The remainder of this paper is structured as follows. We first discuss the
%reading and conversion of values from reduced precision formats. We present
%some experimental results from benchmarks which perform either exclusively read
%accesses or where the large majority of accesses are reads. We then discuss the
%writing of values in reduced precision storage formats, and present some
%experimental results from benchmarks which perform write accesses. Finally, we
%discuss some related work and the significance of our results in the particular
%benchmarking scenario we have chosen.

\section{Access in Reduced Precision}
\label{sec:rw}

Reading and writing reduced precision representations might be expected to
incur a significant performance penalty due to the overheads outlined in
Section~\ref{sec:scalar-approach}. Particular concerns are (1) the overhead of
conversion between data formats, in addition to (2) the overheads of memory
access to arrays of datatypes that may have non--power-of-two byte width, where
memory movements may be overlapping and misaligned. Although these overheads
are encountered both when reading and when writing reduced-precision
representations, there are important differences between the two cases.

\subsection{Reading In Reduced Precision}
\label{sec:reads}

Modern instruction set architectures typically have native support for data
movement using types with power-of-two byte widths -- usually 1, 2, 4, and 8
bytes. Since our \textit{flyte} types differ in width from standard IEEE-754
types by multiples of 8 bits, this means we can always fetch a \textit{flyte}
with a single read using a wider native type (e.g. a 4-byte read for a
\mt{flyte24}).

Since we propose to store \textit{flyte}s packed
consecutively in arrays without padding, the majority of such accesses will be
misaligned. Specifically, a consecutive \textit{flyte} array access will only
be aligned to the next largest native type once every $lcm(
\mathtt{native}_{bits}, \mathtt{flyte}_{bits}) / \mathtt{flyte}_{bits}$ array
elements.

Unaligned access can cause extra cache misses versus aligned access due to the
possibility that the accessed item spans a cache line boundary. The strategy of
using overlapping accesses at the next largest native type allows us to utilize
the vectorization approach of Anderson et al.~\cite{anderson2015automatic}.
(Section~\ref{sec:vector-approach}) for our vector memory accesses. Also, the
conversion process when reading \textit{flytes} is relatively simple, requiring
only that the read data be shifted and padded with zero bits.

\subsection{Writing In Reduced Precision}
\label{sec:writes}

Writing data to a reduced-precision format is more complex than reading, both
in terms of the memory movement (since memory locations are modified), and due
to the fact that when writing, the number format narrows, and precision is
lost.

Loss of precision is a natural consequence of working with floating point
numbers. The precise result of a floating point computation can be a real
number which cannot be exactly encoded in a finite representation (for example,
it might require infinitely many digits). In these cases, loss of precision is
necessary to make the result representable.

The IEEE-754 standard specifies
several methods which are used to \emph{round} numbers to a nearby
representable value. One straightforward way to perform rounding is to simply
truncate extra digits (i.e. bits, in a binary representation). This is the
standard round-to-zero rounding mode~\cite{zuras2008ieee}. Other rounding modes
are specified by the IEEE-754 standard, including round-to-nearest-even and
round-to-infinity (positive or negative).

%In Section~\ref{sec:whentoround} we discuss some
%mechanisms which the programmer can use to control (1) when loss of precision
%occurs, and (2) the method used to convert values to narrower representations.

%In this
%section, we first discuss the vectorization of the data movement, and then the
%task of data conversion, which chiefly involves the issue of \emph{rounding} of
%values from larger to smaller number formats.

\subsection{Vectorized Reading and Writing}
\label{sec:vector-approach}

We propose a compiler-based approach which can greatly reduce the overhead of
operating on \textit{flyte} arrays using automatic vectorization. Our approach
uses vector instructions to load, convert, and store \textit{flytes}. We
generate vectorized code to load a number of \textit{flyte} values at once, and
unpack those values in parallel to the next largest IEEE type using vector
shuffle and blend instructions. We use vectorization not only to amortize the
cost of converting each element (as it does for other operations), but also to
help overcome penalties associated with \texttt{flyte} types' unnaturally
aligned memory accesses.

By restricting the size of a \textit{flyte} to be a multiple of 8 bits, we
ensure that widely available fast byte-level vector reorganization instructions
can be used. Finally, when computation is complete, we again use vector
instructions to convert back to \textit{flyte} types. This may involve a
rounding step when reducing the precision of a standard IEEE-754 format to a
\textit{flyte} type, followed by packing the data in vector registers before
the results are written to memory.

%A vectorized floating point computation may require \mt{VF} \textit{flyte}
%elements to be loaded into or stored from a single vector register. As
%previously outlined, these quantities may not correspond to any native type,
%but must still be loaded from memory, converted, and packed consecutively in
%the lanes of a vector register. Similarly, after computation has produced a
%vector containing \mt{VF} packed results, they must be rounded to the
%appropriate format and packed consecutively for storage.

\begin{figure}[ht]
\centering\includegraphics[scale=1.2]{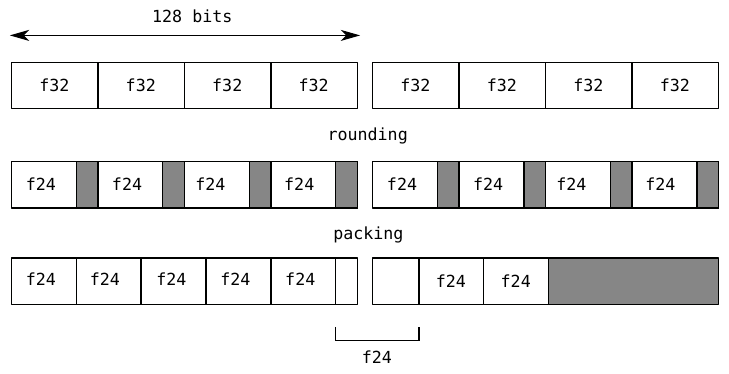}

\caption{Layout of data in 128-bit (16-byte) vector registers. (top) before
format conversion, (center) after rounding to the desired size, and (bottom)
the desired layout in memory. Note that the desired memory layout requires data
elements to straddle the boundary between vector registers.}

\label{fig:f24}
\end{figure}

Vectorized loading and storing of packed data elements that do not correspond
to any native machine type presents additional challenges over scalar memory
movement. Since vector lengths on modern GPPs are usually a power-of-two bytes,
vectorized access to \textit{flyte} arrays often leads to the splitting of data
elements across vector memory operations, where the leading bytes of an element
are accessed by one operation, and the trailing bytes by the next.
Figure~\ref{fig:f24} displays one such scenario: storing in \mt{flyte24} format
computational results produced in \mt{binary32}.

%Typically,
%loads can be performed as normal, using the next largest native type, followed
%by the discarding of some loaded data, and the mapping of the remaining data to
%a legal value in the representation being used for computation. However,
%storing is a more difficult issue. Since stored values do not correspond to
%native machine types, we cannot simply use native store instructions and
%reorganize later, as we can with loads, but must first \emph{pack} the data to
%be stored into native types.

%When vector instructions are used to load and legalize several values at once,
%the cost of the reorganization is amortized across all of the loaded values.

\noindent While vectorized reading is not significantly more complicated than
scalar reading, vectorized writing has additional issues. A straightforward
approach to vectorized writing of unpadded \textit{flyte} data could pack as
many consecutive \textit{flyte} elements in a vector register as would fit, and
perform a store with the trailing unused byte positions predicated off.

Subsequent stores, if there are more than one, could overlap prior stores in
these unused positions, so that the data is consecutive in memory. Due
to the structure of load/store hardware in GPPs, this approach is likely to be
extremely inefficient.

Our vectorized approach to storing values in reduced precision format
works by packing all the rounded data elements to be stored consecutively in a
set of vector registers, which are mapped to a set of consecutive,
\textbf{non-overlapping} vector memory movements (shown in
Figure~\ref{fig:f24}).

We use a two-phase permute and blend approach. Vector
permute instructions are initially used to compact the rounded data in each
register into memory order, and align the contents of some registers so that
the leading data element in each register is located at the position of the
first unused byte in the previous register. Next, the compacted vector
registers are combined together using vector blend instructions until a number
of fully packed vector registers result.

The resulting registers can be stored without overlap, and data elements are
correctly split across register boundaries. If the data written cannot be
packed perfectly into full vector registers, some vector stores may be partial
stores with some additional implementation concerns. These are described in
detail in the vectorization approach of Anderson et
al.~\cite{anderson2015automatic}.

\subsection{Controlling Format Conversion}
\label{sec:whentoround}

Programming language designers have three conflicting goals when deciding the
rules for evaluating floating point expressions. Evaluating expressions at
higher precision may result in more accurate answers, which suggests that
higher precision should be used if it is available.

On the other hand, programmers like to get bit-exact identical results from
their program regardless of which compiler is used, which suggests that the
language should strictly define the precise precision and each floating point
operation.

The IEEE-754 standard stipulates~\cite[\S 11]{zuras2008ieee} that conforming
languages \emph{should} support reproducible programming, and defines
circumstances when programs should have numerically identical results across
compliant platforms. Finally, giving the compiler the freedom to choose
precision may result in more efficient code.

Controlling exactly when format conversion occurs is an important part of using
non-standard formats such as our \textit{flyte}s. Languages such as
C99~\cite{british2003c} provides features aimed at providing consistency in the
output of the same program on different platforms.

In particular, the programmer can choose to specify the rules determining the
precision of intermediate values in expressions using the
\mt{FLT\_EVAL\_METHOD} facility.

The programmer can also use the pragma directive \mt{STDC FP\_CONTRACT} to
enable or disable \emph{contraction} -- atomic evaluation of floating point
expressions which can use higher precision and omit rounding errors implied by
the source code and \mt{FLT\_EVAL\_METHOD} \cite[\S 6.5]{british2003c}. Both
facilities modify behaviour at the expression level.

For example, the programmer can choose to have all subexpressions within an
expression tree evaluated at the exact width of the widest operand to each
operator (by setting \mt{FLT\_EVAL\_METHOD = 0}).

Such strict constraints on the types of all intermediate floating point values
can result in very poor performance for \textit{flytes}. For example,
\mt{flyte24} is stored in memory as a 24-bit value, but we must convert it to
\mt{binary32} before performing any operations. In a larger expression of
\mt{flyte24} values, the generated code is likely to be much faster if all
intermediate values can be kept in \mt{binary32} rather than being converted
down to \mt{flyte24} after every operation.

C99 also provides a feature that allows floating point operations to be
performed at a higher precision than the source values or result:
\emph{contraction} of floating point expressions, which is controlled by the
standard pragma directive \texttt{FP\_CONTRACT}. In all our experiments,
\texttt{FP\_CONTRACT} is on.

One problem not addressed by C99's floating point support is the issue of using
higher precision values across \emph{statements}. C99 requires that the value
stored in a variable must be convered to the type of that variable at the point
where it is written to storage.

We propose a type qualifier \mt{AT\_LEAST}
which is used to tag a floating point type informing the compiler that it is
free to use a higher precision to store results of that type, rather than
converting strictly to the precision of the storage type. This allows the use
of higher-precision types in code like that in Figure~\ref{fig:at-least-demo}
where accumulation into a variable would otherwise result in many lossy
conversions.

\begin{figure}[h]
\begin{verbatim}
flyte24 sum(flyte24 * a, int size)
{
  AT_LEAST flyte24 sum = 0.0;
  for(int i = 0; i < size; i++) {
    sum = sum + a[i];
    /*
       without AT_LEAST, C99 will truncate
       sum here in every loop iteration
     */
  }
  return sum;
}
\end{verbatim}
\caption{Example of the use of the \mt{AT\_LEAST} qualifier to allow accumulation in
a higher precision than the input/output.}
\label{fig:at-least-demo}
\end{figure}

\noindent The code in Figure~\ref{fig:at-least-demo} shows the variable
\mt{sum} of type \mt{AT\_LEAST flyte24} which the compiler can represent as any
floating point type with at least the precision of a \mt{flyte24}. In general,
the next largest native type is a good choice for a use of the \mt{AT\_LEAST}
qualifier.

\section{Rounding}
\label{sec:rounding}

When numbers represented in IEEE-754 floating point format are used in
computations (such as addition and subtraction), the natural result of
computation is often a real number which is not exactly representable in the
finite representation. In these cases, the standard specifies a way to round
these numbers to a nearby representable value.

Computations on \textit{flyte}s are performed using the next largest
standard floating point size.  After each operation, the built-in
floating point type performs its own rounding. However, a question
arises when we convert from IEEE floating point types to
\textit{flyte}s: should we round again during conversion?

Double rounding is discussed in considerable depth by Boldo and
Melquiond~\cite{boldo2004double}, and we do not reproduce their arguments here.
Note that our aim in this paper is simply to demonstrate that a low-overhead
implementation of customized floating point types is feasible on general
purpose processors.

Any compiler framework which might implement our proposed
scheme should take the necessary steps to ensure that the transformations
applied to convert between floating point representations are correct; in the
remainder of this section, we describe some low-overhead mechanisms which can
be used to implement them.

\subsection{Round-towards-zero}
\label{fig:round-towards-zero}
The simplest approach is to round by truncating the lower matissa
bits, an option which is known as \textit{round-towards-zero} in IEEE
754. Rounding towards zero is simple to implement, but IEEE floating
point has a number of special cases that we must treat correctly.

In IEEE-754 a NaN value has an exponent consisting entirely of ones, and a
mantissa value that is non-zero. If the non-zero part of the mantissa is in the
lower bits, truncating those bits may cause the entire mantissa to have a zero
value.  This would change the value from NaN to a value with all ones in the
exponent and zero matissa, which represents infinity in IEEE-754.

However, there are two types of NaNs: signalling NaNs, which cause a floating
point exception, and quiet NaNs which indicate an invalid value with causing an
exception. In IEEE-754 binary floating point formats, quiet NaNs are
distinguished from signalling NaNs by the value of the \emph{most significant
bit} of the mantissa, which is preserved by truncation of up to $M-1$ bits.

In IEEE-754, a subnormal number has a zero exponent, and the mantissa
represents a very small fixed-point number. Truncating the final bits of a
sub-normal number may cause its value to change to zero. This is correct
behaviour, since the non-zero part of the number is too small to represent in
our smaller \textit{flyte} type, and zero is the closest representable value.

\subsection{Round-to-nearest}

Round-towards-zero is simple to implement, but it can result in large errors.
For example, if round-towards-zero were applied in decimal, the number 9.9
would be rounded to 9, rather than 10. The maximum error can be reduced by
rounding to the nearest representable number rather than simply truncating.

A special case is where the number to be rounded is exactly between two values.
To break the tie, IEEE-754 specifies that exact ties should be rounded to the
nearest even number.

\begin{figure}[h]
\centering\includegraphics[scale=0.8]{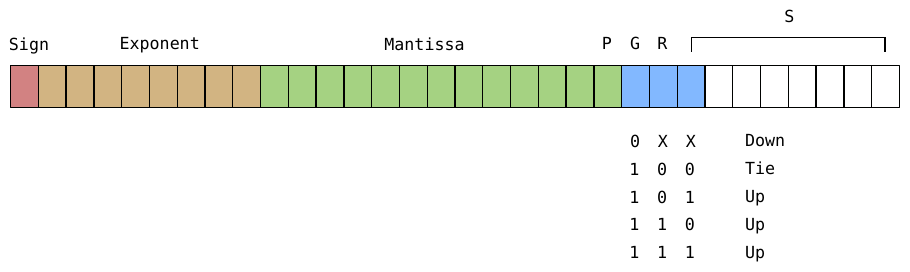}
\caption{Rounding to nearest even. Marked bit positions correspond to
\textbf{G}uard, \textbf{R}ound, and \textbf{S}ticky bits. To avoid
overflow into the exponent when rounding to nearest even in
software, the pre-guard bit (\textbf{P}) must also be inspected.}
\label{fig:rounding}
\end{figure}

Figure~\ref{fig:rounding} shows a 32-bit floating point number being rounded to
our 24-bit representation using round-to-nearest, where exact ties between two
values are rounded to the nearest even value.

Rounding to nearest even is expressed in terms of the relationship between
three bits (Guard, Round and Sticky) around the point where the number is
rounded. There are no hardware instructions in conventional processors for
rounding floating point values to \textit{flytes}, and we must therefore round
in software.

\noindent In the most straightforward case, rounding to nearest even simply
involves computing the nearest representable number. An easy way to do this is
shown in Figure~\ref{fig:rounding-to-nearest}.

\begin{figure}[h]
\centering
\begin{verbatim}
flyte24 round_to_nearest(float num)
{
  u32 temp = cast_f32_to_u32(num);
  // round by adding 0.5 ULP
  temp = temp + 128;
  // truncate last eight bits
  temp = temp >> 8;
  return cast_u32_to_flyte24(temp);
}
\end{verbatim}
\caption{Heuristically rounding to nearest even by adding half of a unit of least precision (ULP).}
\label{fig:rounding-to-nearest}
\end{figure}

\noindent The code in Figure~\ref{fig:rounding-to-nearest} adds half of a unit
of least precision (ULP) in the new smaller format to the value in the existing
larger format.

As long is the number to be rounded is not exactly between two representable
values in the new format, this will result in correct round-to-nearest-even.
Implementing precise round-to-nearest-even requires checking for this tied
special case, and rounding accordingly. In addition, there are numerous special
cases which must be checked to correctly implement IEEE-754 mandated behaviour.

\subsection{Treatment of Special Values}

The IEEE-754 floating point standard has special values and ranges as
previously outlined in Section~\ref{sec:ieee-754-intro}. Rounding interacts in
different ways with these values and ranges, and behaviour which may be
appropriate for some scenarios may not be for others.

The application programmer must make a choice of rounding approach based on the
information available. We describe the behaviour of each of our proposed
rounding approaches here with respect to IEEE-754 special values and ranges.

\subsubsection{Normalized numbers}

When a normalized number is being rounded, an infinity occurs when the
rounded value is so large that the exponent is all ones after rounding
(overflow). This is a natural consequence of conversion from a larger to a
smaller finite representation.

However, when a normalized number is so small that its exponent is all zeros
after rounding (underflow), it does not get rounded directly to zero, but
instead to a subnormal number.

Underflow is \emph{gradual}, and a number will underflow to zero only when it
is so small that both exponent and mantissa are all zeros after rounding.
Normalized numbers may therefore naturally be rounded to several different
classes of value.

Very large positive or negative numbers may go to infinity when rounded, and
very small numbers may become subnormal. This behaviour conforms to IEEE-754
semantics.

\subsubsection{Subnormal numbers}

Subnormal numbers are distinguished by a zero exponent, and lack the implied
leading 1 in their mantissa. They represent numbers very close to zero. The
expected behaviour of format conversion for subnormal numbers is slightly
complicated due to the issues of overflow and underflow.

The closest value in the target representation for a very large subnormal
number may be a very small normalized number (overflow), while the closest
value for a very small subnormal number may be zero (underflow).

For subnormal numbers, rounding may validly cause the class of the value to
change, either by underflow to zero, or by overflow to a small normalized
number.

\eject
\subsubsection{Infinities}

Positive and negative infinities are encoded with an exponent which is all 1s
and a zero mantissa. There are only two values in this class, which are
distinguished from each other by their sign. The expected behaviour of format
conversion for infinities is a correctly signed infinity in the target format.

\subsubsection{NaN values}

NaN values represent the result of expressions of \emph{indeterminate form},
which cannot be computed, such as $\infty-\infty$. The expected behaviour of
format conversion of a NaN value is a NaN value in the target format. However,
NaN is not a singular value, but a value \emph{range}. NaN values occupy a
range of bit patterns distinguished by an exponent which is all ones and any
nonzero value in the mantissa.

Since the mantissa is truncated by conversion to a shorter format,
some NaN values cannot be represented after down-conversion, and
indeed truncation may cause the non-zero part of the mantissa to be
lost entirely.

However, as described in section \ref{fig:round-towards-zero}, IEEE-754 non
signalling (or \textit{quiet}) NaNs always have a 1 in the \emph{most
significant bit} of their mantissa. Thus, although the exact mantissa value of
a NaN may change after truncation, a quiet NaN cannot become a non-NaN value
through truncation.

Signalling NaNs are different: the non-zero bits of a signalling NaN's
mantissa may be entirely in the lower bits. A signalling NaN may be
corrupted to become an infinity value by truncation. Thus, signalling
NaNs should not be used with \textit{flyte} values, unless the signal
handler is modified to place a non-zero value in the higher-order bits
of the NaN.

When using our add-half-an-ULP heuristic, a further problem can arise with
non-signalling NaNs. If the mantissa value of the non-signalling NaN is the
maximum representable value after truncation, then adding even half an ULP will
cause an overflow from the mantissa to the exponent, and potentially into the
sign bit, if the exponent is all ones.

Our slowest and most correct round-to-nearest mode checks for this case, and
corrects the value if necessary. Our fast heuristic round to nearest approach
(Figure~\ref{fig:rounding-to-nearest}) does not. However, despite extensive
testing we have never seen a case where the floating point unit creates such a
pathological NaN value as the result of arithmetic.

\section{Experimental Evaluation}
\label{sec:eval}

We benchmarked the performance of our proposed scalar code implementation from
Section~\ref{sec:scalar-approach} and vectorized implementation from
Section~\ref{sec:vector-approach}. Experiments were run on 64-bit Linux with a
4.2 series kernel, using a machine with 16GB of RAM and an Intel Core i5-3450
(Ivy Bridge) processor.

\begin{figure*}[ht]
\centering

\subfloat[Scalar code (Approach shown in Figure~\ref{fig:simple-flyte})]{
\includegraphics[scale=0.75]{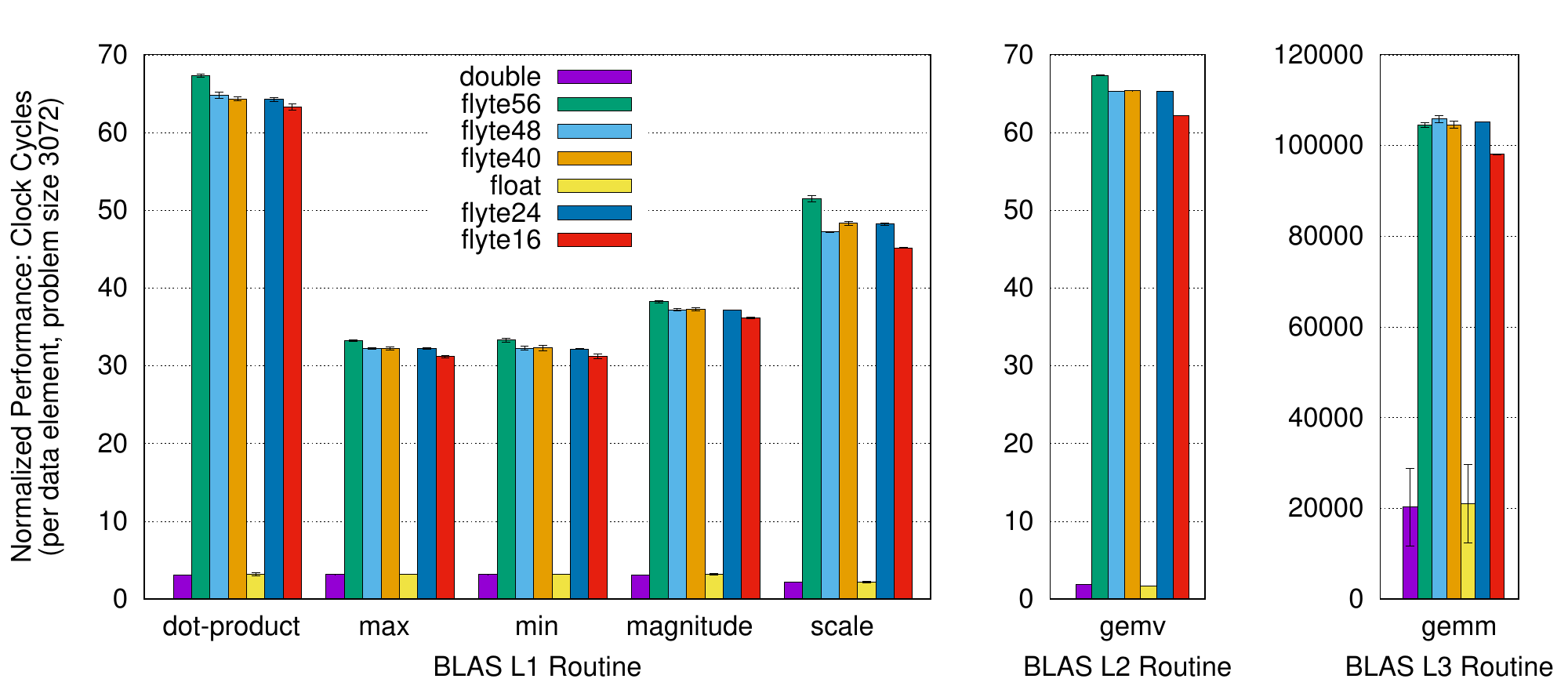}
\label{fig:realworldlib}
}\qquad
\subfloat[128-bit SSE code (Approach shown in Figure~\ref{fig:f24})]{
\includegraphics[scale=0.75]{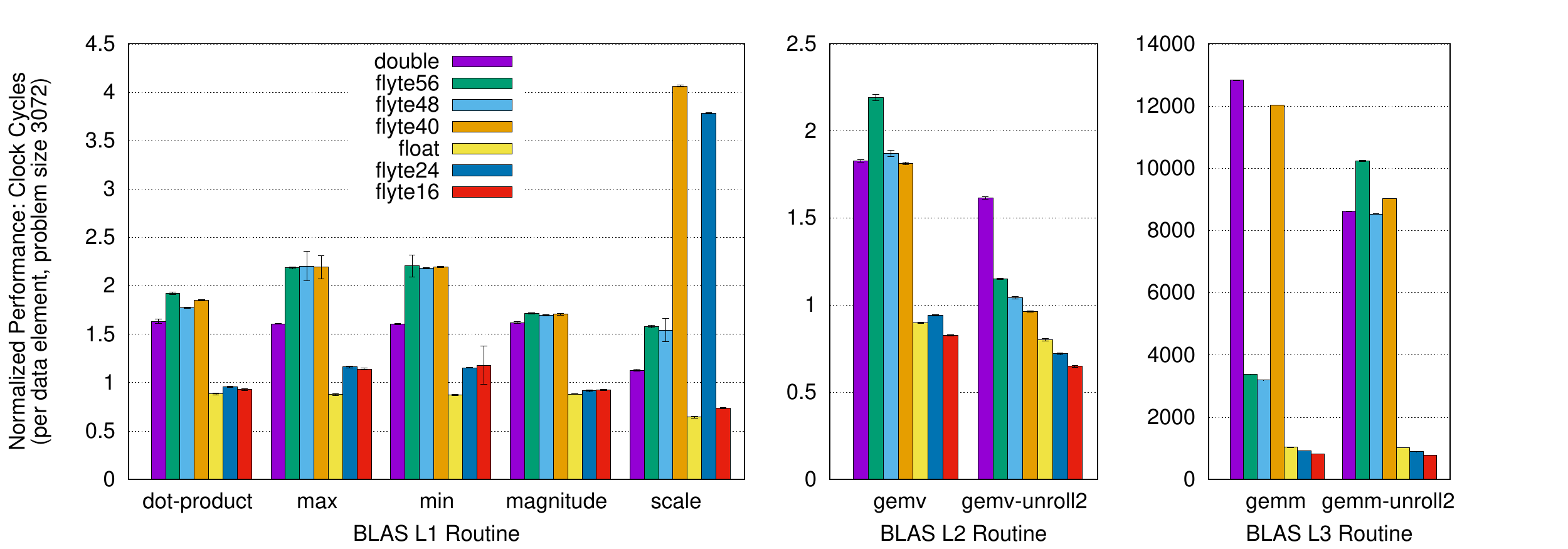}
\label{fig:realworldsse}
}

\caption{Variation of performance with precision of memory representation
across a number of BLAS kernels. Performance displayed as normalized execution
time (cycles per data element) -- \textbf{lower is better}. Overhead versus
native types can be read as the difference between any \mt{flyte} type and the
next largest native type.}

\label{fig:realworld}

\end{figure*}

We followed Intel's guidelines for benchmarking short programs on this
architecture~\cite{paoloni2010benchmark}. Figures~\ref{fig:realworld}
and~\ref{fig:overhead} present the results of benchmarking. In our experiments,
we use rounding towards zero, \texttt{FP\_CONTRACT} is on, and variables which
are used to accumulate are declared as \texttt{AT\_LEAST} the target precision.

Problem size in experimental figures refers to the number of elements in arrays
in BLAS operations - for vector-vector operations (BLAS Level 1) this is the
number of elements in a vector, while for operations involving matrices (BLAS
Level 2 and 3) this is the number of elements in a row or column of a square
matrix, so that the total number of data elements is the square of the problem
size.

\subsection{Overheads: SIMD vs Non-SIMD} The large overhead of scalar access in
Figure~\ref{fig:realworldlib} is due to the design mismatch between our
non-standard use-case and the typical structure of a GPP scalar datapath,
discussed in more detail in Section~\ref{sec:scalar-approach}. The GPP datapath
is ill equipped to handle simultaneous misaligned accesses to data stored in
packed non--power-of-two multibyte formats.

In contrast, our compiler-based vectorized approach
(Figure~\ref{fig:realworldsse}) marshalls and unmarshalls this data into
consecutive, non-overlapping power-of-two length accesses which are the
best-case for performance using the (SSE) vector datapath.

Overheads in the scalar implementation are in the mid to high tens of cycles
per accessed element, while in the vectorized implementation, overheads versus
native IEEE-754 types are on the order of a single cycle per data element. In
computationally heavy programs like \texttt{magnitude} this overhead is
effectively hidden by instruction-level parallelism
(Figure~\ref{fig:realworldsse}).

The relatively high overhead of \mt{flyte40} and \mt{flyte24} in the
\texttt{scale} benchmark in Figure~\ref{fig:realworldsse} is due to two
factors: the alignment of the accesses is odd (5 and 3 bytes, respectively)
meaning that data must be shuffled \emph{between} vectors, rather than simply
within vectors, as with other types. Furthermore, the benchmark performs an
\emph{in-place} update of the data, where reads and writes overlap, which
reduces the available instruction-level parallelism.

\begin{figure*}[!ht]
\centering

\subfloat[Normalized variation in absolute performance (clock cycles per element
processed) at each BLAS level as problem size
increases. \textbf{Lower is better}.]{
\includegraphics[scale=0.47]{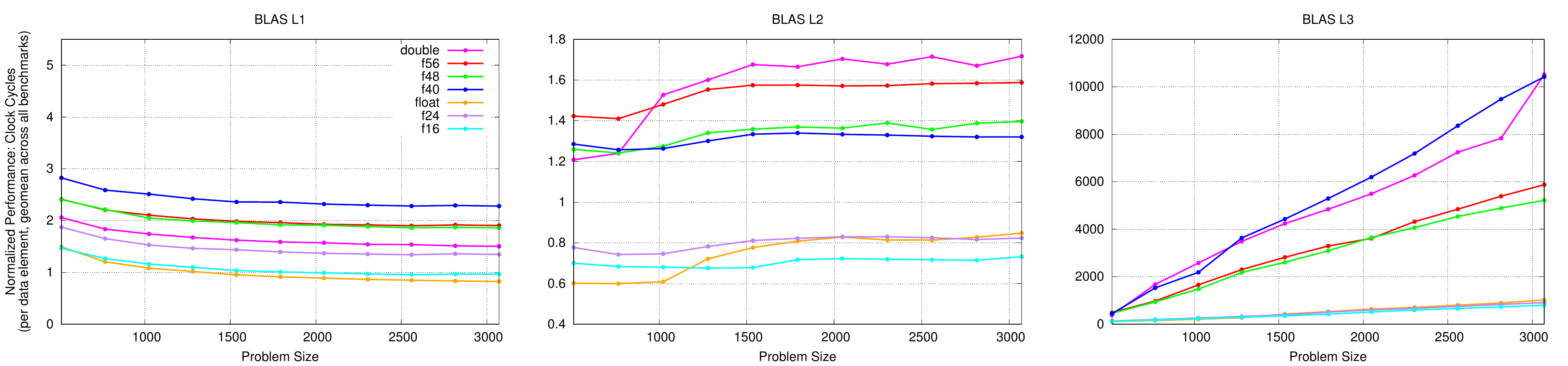}
\label{fig:realworldsse-cyc}
}\qquad
\subfloat[Normalized variation in second-last level cache misses as problem size
increases (misses per element processed). \textbf{Lower is better}.]{
\includegraphics[scale=0.47]{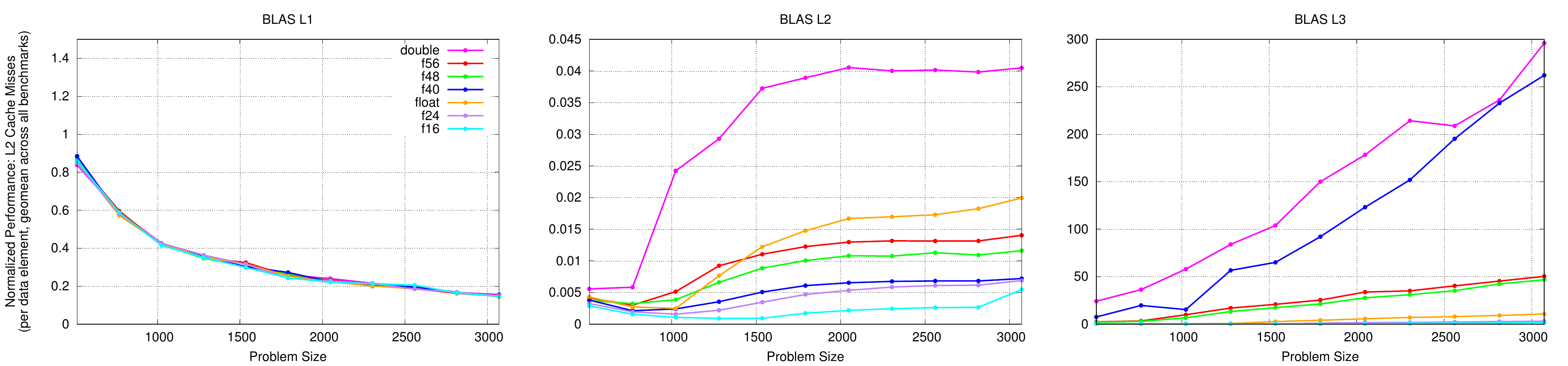}
\label{fig:realworldsse-l2m}
}\qquad
\subfloat[Normalized variation in last-level cache misses as problem size
increases (misses per element processed). \textbf{Lower is better}.]{
\includegraphics[scale=0.47]{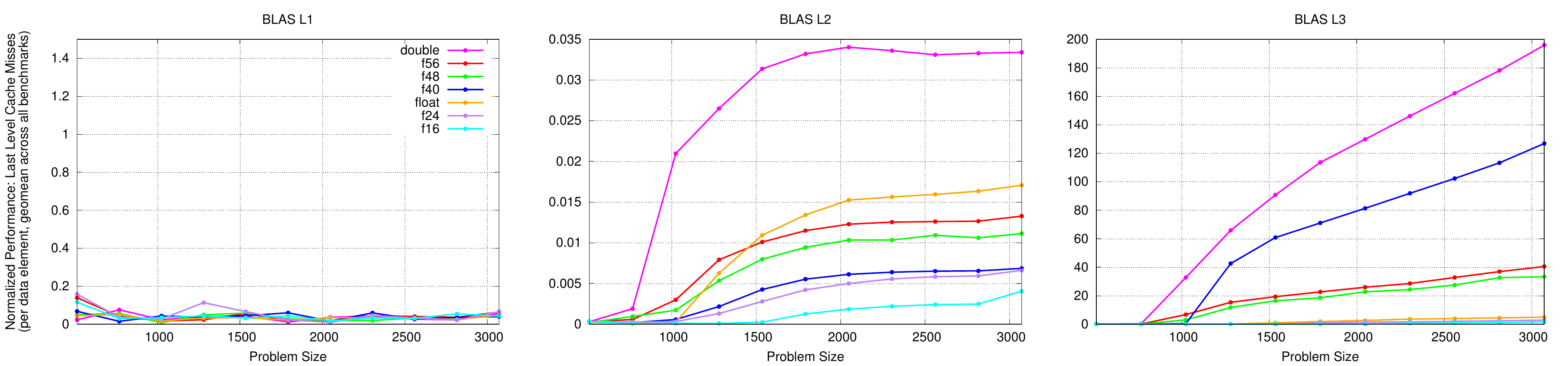}
\label{fig:realworldsse-l3m}
}

\caption{Summary of absolute performance (cycles per processed element) and
cache behaviour (cache misses at L2 and L3) for our vectorized BLAS benchmark
programs as problem size increases. Performance for each BLAS level is measured
as the geometric mean performance across all the programs in that level. The
programs in each BLAS level are shown in Figure~\ref{fig:realworldsse}.}

\label{fig:overhead}

\end{figure*}

However, the overhead is still on the order of 3 cycles per data element in the
worst case, which may be perfectly acceptable in many scenarios in return for a
37.5\% reduction in memory traffic. Indeed, as can be seen from our results in
Figure~\ref{fig:realworldsse-cyc}, \texttt{flyte40} is only marginally slower overall
considering BLAS Level 1 programs, and significantly reduces second-last and
last-level cache misses for BLAS Level 2 and Level 3 programs.

%TODO: REWRITE THIS BIT
%Our results show that reducing memory traffic with non-standard floating point
%memory representations can increase performance. However, the picture is much
%better for our compiler-based vectorized approach. Where available memory is a
%pressing concern, the tradeoff of a 37.5\% reduction in memory requirements for
%3 cycles overhead per accessed element may even be attractive.

\subsection{Effect of Unrolling Loops}

The benchmark \texttt{gemv-unroll2} in Figure~\ref{fig:realworldsse} is the
BLAS Level 2 \texttt{GEMV} kernel unrolled twice to increase the amount of data
movement per vectorized loop iteration. In this benchmark, data transfer
accounts for a large portion of total execution time.

The benchmark demonstrates that a win-win is possible: the choice of a
reduced-precision memory representation can actually \emph{increase} overall
performance versus the next largest IEEE-754 type, while also reducing memory
requirements, even on a general purpose processor without special hardware
support for non-standard floating point.

The benchmark \texttt{gemm-unroll2} in Figure~\ref{fig:realworldsse} is the
BLAS Level 3 \texttt{GEMM} kernel unrolled twice to increase the amount of data
movement per vectorized loop iteration.

In effect, unrolling is equivalent to a simple 1-dimensional loop tiling with
tile size $2\times\mathtt{VF}$ where \mt{VF} is the vectorization factor. In
this benchmark, data transfer accounts for a large portion of total execution
time.

However, unlike \texttt{GEMV}, \texttt{GEMM} exhibits a slowdown when unrolled.
We inspected the performance data and found that the number of last level cache
misses was significantly elevated when \texttt{GEMM} was unrolled.

In this case, unrolling simply introduces too much data movement in the inner
loop. It is likely that less na\"ive tilings could mitigate this effect,
however, we aim only to show the effect of using reduced precision types.

\eject
\subsection{Effect on Cache Behaviour}

The primary effect of using shorter data representations is seen in the
behaviour of the second-last and last-level caches. Using a shorter data
representation means that each individual memory operation (i.e. cache line
read or write from DRAM) stores or retrieves a larger proportion of data
elements, resulting in fewer cache misses overall. Reducing the number of
last-level cache misses in particular has a large effect on performance, and on
energy efficiency~\cite{gustafson}.

\subsubsection{BLAS Level 1} Figure~\ref{fig:overhead} displays a summary of
the absolute performance (cycles per data element) as well as the second-last
and last-level cache behaviour for our benchmark programs. As in
Figure~\ref{fig:realworld}, the programs are divided into three categories, one
for each BLAS level.

For BLAS level 1 programs, we see that the variation in absolute performance
remains small as problem size increases. BLAS level 1 programs are mostly
compute-bound, so using smaller types does not significantly affect
performance.

Figures~\ref{fig:realworldsse-l2m} and~\ref{fig:realworldsse-l3m} show that,
for BLAS Level 1, there is very little difference in the number of second-last
and last-level cache misses from using shorter types.

\subsubsection{BLAS Level 2} For BLAS level 2 programs, we see that initially
there is little variation in performance at small problem sizes. However, as
the problem size increases, the effect of using smaller types becomes apparent.
For BLAS Level 2 programs, we see a large variation in performance once the
problem size exceeds the capacity of the L1 cache.

In Figure~\ref{fig:realworldsse-cyc}, we see that for BLAS Level 2 programs,
\texttt{float} and \texttt{double} initially outperform their reduced-precision
representations, but once the problem size grows large enough, the situation is
reversed.

Moreover, smaller representations outperform larger ones, in general. At the
largest problem size in experiments, \texttt{GEMV} on \texttt{flyte40}
outperforms \texttt{GEMV} on \texttt{double} by 23.5\%
(Figure~\ref{fig:realworldsse-cyc}, center graph).

Figures~\ref{fig:realworldsse-l2m} and~\ref{fig:realworldsse-l3m} display the
large reduction in second-last and last-level cache misses for BLAS Level 2
programs. In some cases, the reduction is as many as $6\times$ fewer last-level
cache misses (compare \texttt{flyte40} and \texttt{double} in
Figure~\ref{fig:realworldsse-l3m}, center graph).

\subsubsection{BLAS Level 3} For BLAS Level 3, we again see that using smaller
types results in many fewer second-last and last-level cache misses, although
the reduction is less pronounced for \texttt{flyte40}. 5-byte accesses are
frequently split across cache lines, causing two misses, which offsets the
reduction in misses from simply transferring less data overall.

However, we again see that performance closely tracks cache behaviour - our
straightforward implementation of \texttt{GEMM} is heavily memory-bound, so
this is not surprising. Overall, we see a significant reduction in cache misses
for smaller types: as many as $4.5\times$ fewer last-level cache misses
comparing \texttt{flyte56} and \texttt{double}
(Figure~\ref{fig:realworldsse-l3m}, rightmost graph).

\section{Related Work}
\label{sec:related}

Much prior work discusses reduced precision floating
point~\cite{rubio2013precimonious,lam2013automatically,buttari2006exploiting}.
Jenkins et al.~\cite{jenkins2012byte} evaluate a reduced-precision scheme using
GPPs in an extreme-scale computing context. They do not utilize SIMD, address
only reads, and convert in a pre-pass.

Many approaches use FPGAs or otherwise customized hardware; notably Tong et
al.~\cite{tong2000reducing}, who propose customizing ALUs to support
short-mantissa representations. More recently, De Dinechin et
al.~\cite{de2009generating} also propose custom hardware to support reduced
precision. Ou et al. accelerate mixed-precision floating point using a vector
processor with a customized datapath~\cite{ou2015mixed}.

Our approach, since it targets GPPs, is necessarily less flexible than
FPGA/custom hardware based approaches. However, it is precisely because GPPs
are so widely deployed that reduced precision support on GPPs is attractive.

Prior work on loading only portions of floating point numbers by Jenkins et
al.~\cite{jenkins2012byte} does not perform an explicit rounding step, but
directly truncates values. They perform a byte-level transpose on a matrix of
floating point numbers stored in memory, chopping off some number of trailing
bytes of each number.

This conversion approach is semantically equivalent to rounding to zero.
Jenkins et al. evaluate the error introduced by rounding to zero and find it
acceptable for their purposes.

In contrast, using our proposed techniques, the rounding step in
Figure~\ref{fig:f24} can be implemented in any way which is suitable for the
needs of the application, including rounding towards zero, rounding to nearest
even, or rounding to odd as proposed by Boldo and
Melquiond~\cite{boldo2004double}.

\section{Conclusion}
\label{sec:conclusion}

In this paper, we propose \textit{flyte}s; a scheme representing floating-point
data in memory at precisions along a continuum between IEEE-754 types,
converting to and from standard IEEE-754 types to perform computations.

We propose a method for converting between IEEE-754 floating point and
\textit{flyte}s, and show how it can be accelerated using vectorization on
general purpose processors, without requiring special hardware support.

Our proposed technique handles both reads and writes, and supports reduced
precision floating point memory representations with very low overhead.

Our investigation shows that reducing the precision of floating point data in
memory, and using SIMD operations as the basis of a compiler-accelerated scheme
for performing conversions presents a low-overhead path to supporting
customized floating point on commodity general purpose processors.

%to accelerate the and amortize overheads across multiple data
%accesses. Our proposed approach supports the use of fine-grained multibyte data
%formats, and is not limited to the available native machine types. We also
%demonstrate that native hardware capabilities can be used to support format
%conversions on non-native multibyte data formats with very little overhead.

\section*{Acknowledgements}

We would like to thank Ayal Zaks for many helpful comments on an initial draft
of this paper. We would also like to thank the reviewers of PACT 2016 for their
close attention which helped us greatly in improving the presentation.

This work was supported by Science Foundation Ireland grant 12/IA/1381, and
also by Science Foundation Ireland grant 10/CE/I1855 to Lero --- the Irish
Software Research Centre (www.lero.ie).

\eject
\bibliographystyle{abbrv}
\bibliography{main}

\end{document}